\documentclass[twocolumn, times, tighten,twocolappendix, a4paper]{aastex701}
\usepackage{amsmath,amssymb}
\usepackage{txfonts}
\usepackage[T1]{fontenc}
\usepackage{graphicx,multirow,hyperref,url,color,xspace}

\newcommand{\g}{$\gamma$}
\newcommand{\msun}{{M}_{\sun}}

\newcommand{\nustar}{{NuSTAR}\xspace}

\newbox\grsign \setbox\grsign=\hbox{$>$} \newdimen\grdimen \grdimen=\ht\grsign
\newbox\simpropbox
\setbox\simpropbox=\hbox{\raise.5ex\hbox{$\propto$}\llap
 {\lower.5ex\hbox{$\sim$}}}\ht2=\grdimen\dp2=0pt

\begin{document}
\defcitealias{Draghis25b}{D25}
\newcommand{\dr}{{\citetalias{Draghis25b}}\xspace}

\title{Model dependence of XRISM black-hole spin constraints in Cyg X-1}

\author[0000-0002-0333-2452]{Andrzej A. Zdziarski}
\affiliation{Nicolaus Copernicus Astronomical Center, Polish Academy of Sciences, Bartycka 18, PL-00-716 Warszawa, Poland} 
\email[show]{aaz@camk.edu.pl}

\author[0000-0003-3499-9273]{Swadesh Chand}
\affiliation{Institute of Astronomy, National Tsing Hua University, Hsinchu 300044, Taiwan}
\email[show]{swadesh.chand@gmail.com}

\author[0000-0001-7606-5925]{Micha{\l} Szanecki}
\affiliation{Faculty of Physics and Applied Informatics, {\L}{\'o}d{\'z} University, Pomorska 149/153, PL-90-236 {\L}{\'o}d{\'z}, Poland}
\email{mtszanecki@gmail.com}

\author[0000-0003-1589-2075]{Gulab Dewangan}
\affiliation{Inter-University Centre for Astronomy and Astrophysics, Pune 411007, India}
\email{gulabd@iucaa.in}

\author[0000-0003-2743-6632]{Barbara De Marco}
\affiliation{Departament de F\'isica, EEBE, Universitat Polit\`ecnica de Catalunya, Av. Eduard Maristany 16, Barcelona, E-08019, Spain.}
\affiliation{Institut d’Estudis Espacials de Catalunya, C. Esteve Terradas 1, Castelldefels, E-08860, Spain}
\email{barbara.de.marco@upc.edu}

\begin{abstract}
We study the persistent black hole X-ray binary Cyg X-1, recently observed by XRISM Resolve and simultaneously by NICER and NuSTAR in its hard spectral state. We confirm the result of Draghis et al.\ that fits of the Resolve data alone with the simplest available relativistic reflection model, {\tt relxill}, yield a black hole spin parameter close to the maximum, $a_* \approx 0.99$. However, fitting with an improved, Comptonization-based model, {\tt relxillCp}, yields a low $a_*=0.0^{+0.17}$. A similarly low range is obtained with another Comptonization-based model, {\tt reflkerrD}. Then, fits to the combined data require two Comptonization model components but are consistent with any spin value. We conclude that the spin value of Cyg X-1 is strongly model-dependent. However, low spin values are consistent with the constraints from gravitational waves. All of the models constrain the inner disk radius to be $\lesssim$10 gravitational radii, which is consistent with a recent finding of the weakness of thermal reverberation in Cyg X-1. The suggested source geometry is that of an outflowing disk corona, which was also proposed to explain the X-ray polarization observed from this source.
\end{abstract}

\section{Introduction} \label{intro}

The archetypal black hole (BH) X-ray binary (XRB), Cyg X-1, was discovered as an X-ray source in 1964 \citep{Bowyer65}, i.e., more than 60 years ago, and is the best-studied BH XRB to date. Nevertheless, substantial uncertainties in its parameters persist. In particular, according to \citet{Miller-Jones21}, the binary inclination of Cyg X-1 is $i=27\fdg 5_{-0.6}^{+0.8}$, the BH mass is $M=21.2\pm 2.2 \msun$, and the distance to the source is $D= 2.22^{+0.18}_{-0.17}$\,kpc (which we give here as the median values with 68\% uncertainties). By contrast, a recent investigation by \citet{Ramachandran25} using detailed atmospheric non-LTE models of the donor found a significantly lower BH mass, $M\approx (12.7$--$17.8)\msun$. When they used the rotational broadening measurement of \citet{Simon-Diaz17}, they obtained $i\approx 34\degr$ and $M\approx 14\msun$ (see also figure 7 in \citealt{Ramachandran25}). 

Another uncertainty concerns the value of the BH spin parameter, $a_*$. Cyg X-1 has been measured to have very high spin in its soft spectral state (see \citealt{DGK07} for a spectral state definition) using the X-ray continuum method \citep{McClintock14}, e.g., $a_*> 0.9985$ \citep{Zhao21_CygX1, Miller-Jones21}, $0.99964^{+0.00003}_{-0.00007}$ \citep{Steiner24}. Recently, \citet{Draghis25b}, hereafter \dr, used the X-ray reflection method (\citealt{Bambi21} and references therein) to analyze a XRISM \citep{Tashiro21} observation of Cyg X-1 in the hard spectral state, accompanied by a NuSTAR observation \citep{Harrison13}. They also obtained a high spin, $a_*\approx 0.98$. 

However, Cyg X-1 is a high-mass X-ray binary (HMXB) with an OB supergiant donor \citep{Walborn73}. As such, it belongs to the class of potential progenitors of BH mergers \citep{Ramachandran25}. The spins of merging BHs measured by gravitational waves are generally low; the modeled distribution of individual spins peaks at $ a _* \approx 0.01$--0.23, and $\sim$90\% have $a _* \lesssim 0.57$ \citep{LVK25b}. Furthermore, when comparing merger results with XRBs, only the spins of the first-born BHs are of interest. They are plausibly identified with the slower-spinning premerger BHs, while the faster-spinning ones undergo tidal spin-up \citep{Qin18, Mandel20, Ma_Fuller23, Olejak21}. Some higher spins measured in mergers appear to originate from dynamical interactions \citep{Tong25, Antonini2025}, with accretion appearing ineffective at spinning up BHs \citep{vanSon20}. The consensus from those studies is that natal BH spins are low. This finding aligns with stellar models that predict the core remains coupled to the outer envelope during expansion (upon leaving the main sequence), i.e., angular momentum transport is efficient \citep{Spruit02, Fuller_Ma19, Belczynski20}. Those models yield natal spins of $ a _* \sim 0.01$--0.1.  

For Cyg X-1, evolutionary calculations indicate that only hyper-Eddington conservative accretion, at $\dot M \sim 10^{-2}\msun$/yr, could spin up its BH to $a_*> 0.98$ \citep{Qin22}. Those authors invoked the production of neutrino pairs in the accretion flow to remove excess angular momentum, thereby enabling the required mass accretion rate onto the BH. However, the density of such a flow remains orders of magnitude below that required for neutrino pair production, as shown by \citet{Zdziarski24b, Zdziarski26}. On the other hand, accretion rates onto the BH appear to be approximately limited to the Eddington rate, with most of the mass transferred from the donor forming strong winds, e.g., \citet{Fragile25}. 

The same tension applies to two other BH HMXBs with measured spins, LMC X-1 and M33 X-7, whose spins were also found to be high using the continuum method: $a_*=0.92^{+0.05}_{-0.07}$ \citep{Gou09} and $a_*=0.84 \pm 0.05$ \citep{Liu08}, respectively. It remains unclear how to reconcile the gravitational wave results with those from BH HMXBs, a problem reviewed by \citet{Zdziarski26}.

An important caveat for the continuum method is that the theoretical models of accretion disks used in it \citep{NT73, Abramowicz88, Sadowski09} predict that disks are unstable in the radiation-pressure-dominated regime \citep{Lightman74, Shakura76}. This approximately corresponds to the range of Eddington ratios observed in the soft states of BH XRBs. However, observations show that the soft state is stable in X-rays \citep{GD04}. Disks can be stabilized by large-scale magnetic fields, but this substantially alters the structure of the accretion flows, particularly by increasing dissipation in the upper layers of the disk (see \citealt{Zdziarski26} and references therein). This effect has been phenomenologically modeled by adding a warm corona on top of the disk \citep{Belczynski24, Zdziarski24b}. This modeling has resulted in fits to the Cyg X-1 spectra with very low spin values, $a_*\approx 0$. Additionally, the disk's inner edge can move downstream of the innermost stable circular orbit (ISCO), which would also reduce the fitted $a_*$.

There are also caveats regarding the other main method of spin measurement, which is based on X-ray reflection. For this method to be applicable, the disk must extend down to the ISCO, or very close to it. The space-time metric becomes weakly dependent on spin at $\gtrsim 10 R_{\rm g}$, so if the disk is substantially truncated, spin cannot be measured. The method has been applied mostly in the hard spectral state, for which there are strong arguments for truncation (see, e.g., \citealt{DGK07}). While studies by \citet{Duro11}, \citet{Fabian12}, and \dr show disks extending close to the ISCO of a maximally rotating BH, those by \citet{Makishima08}, \citet{Yamada13}, and \citet{Basak17} show substantial disk truncation in Cyg X-1. A related important issue concerns the shape of the incident spectrum, which likely consists of multiple spectral components and thus strongly affects the measurements (see, e.g., \citealt{Zdziarski21b}).

Here, we revisit the Cyg X-1 data set associated with the XRISM observation. We spectrally fit the XRISM spectrum, as well as the NICER \citep{Gendreau16} and NuSTAR \citep{Harrison13} spectra from observations conducted during the XRISM observation. 

\section{The data, lightcurves and variability}
\label{data}

We use the same XRISM observation as in \dr. Cyg X-1 was observed over half of its orbital period, at the phases \citep{Gies08} of 0.67--1.15. Absorption by the stellar wind increased strongly at orbital phases $\gtrsim$0.95. \dr split the observation into halves and analyzed them separately. In the present work, we aim to avoid complications associated with strong absorption while maximizing exposure. We thus choose, for our analysis, the initial 42 h of the observation, excluding the strongly absorbed portion. 

\begin{figure}
\centerline{\includegraphics[width=\columnwidth]{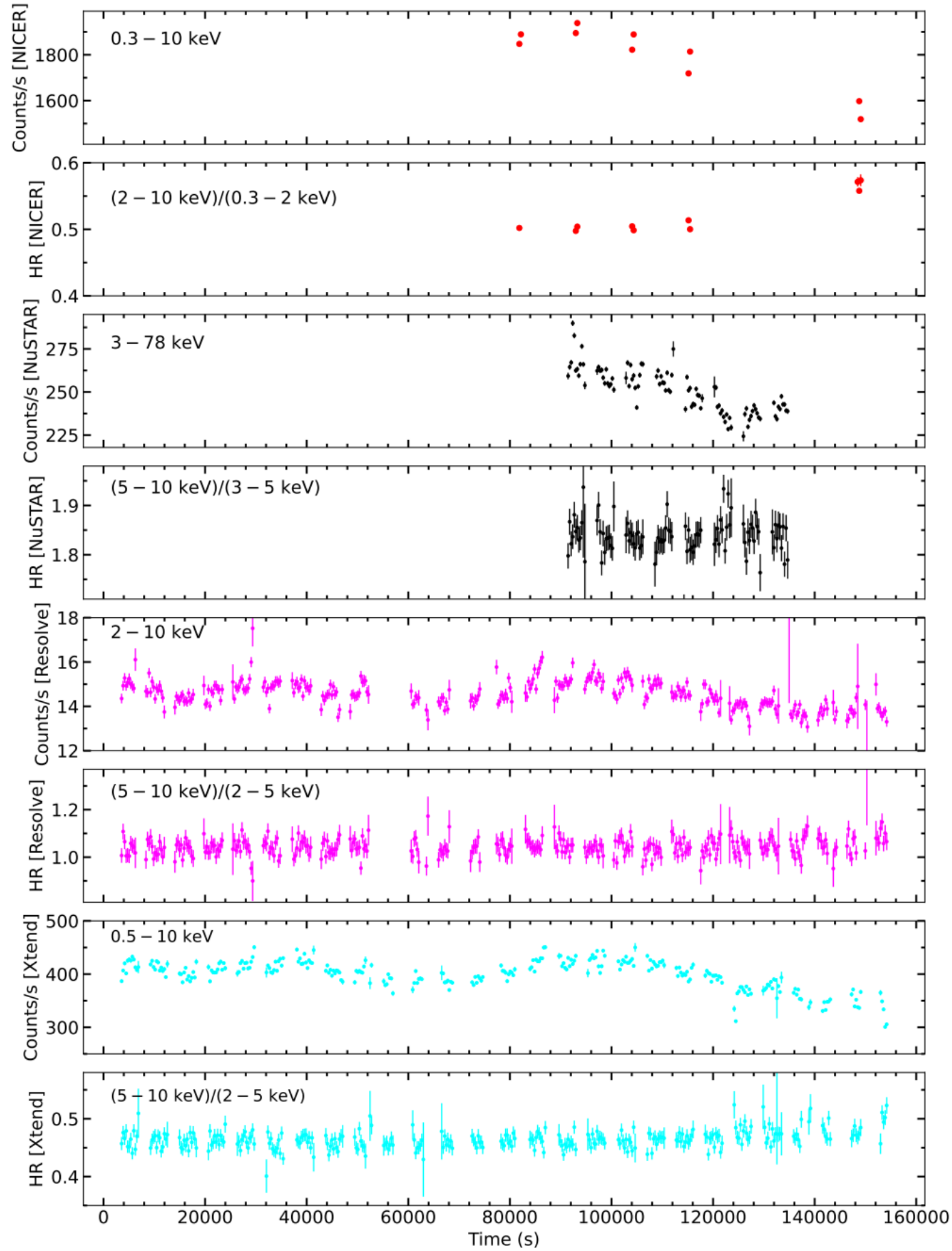}}
 \caption{The light curves from NICER, NuSTAR, XRISM Resolve and Xtend (from top to bottom), together with the corresponding hardness ratios, for the first 42 h studied in this work. The zero time is set to the start time of the Resolve observation, i.e., 2024-04-07 17:56:58. 
}\label{lc}
\end{figure}

\begin{figure*}
\centerline{\includegraphics[width=8cm]{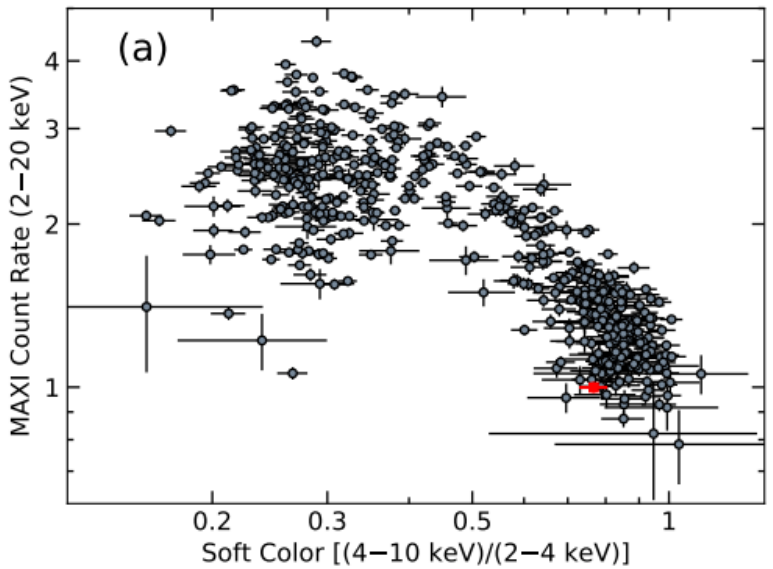}
\includegraphics[width=8cm]{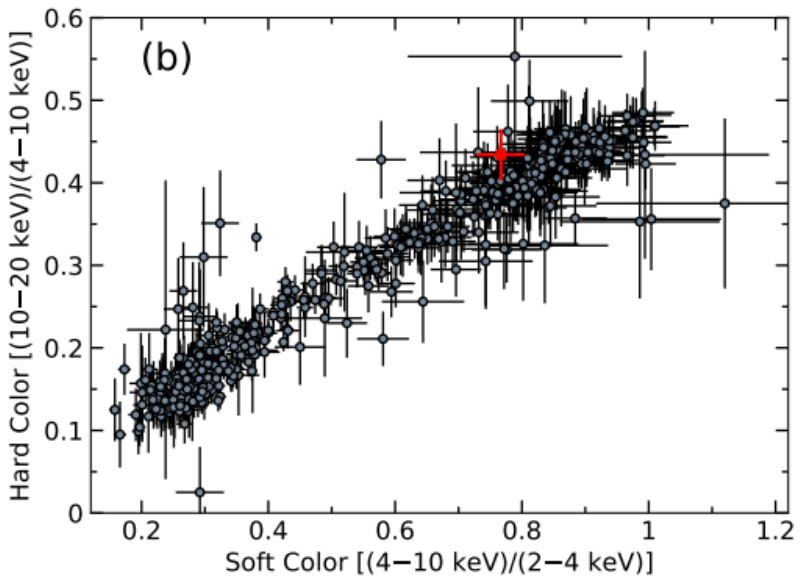}}
 \caption{(a) The 2--20 keV count rate vs.\ the (4--10)/(2--4) keV hardness ratio. (b) The (10--20)/(4--10) keV hardness ratio vs.\ the (4--10)/(2--4) keV hardness ratio (the color-color diagram). Both panels show MAXI data binned over 10-day intervals. The MAXI data averaged over the days of the XRISM observation we used are shown as red symbols.
}\label{hr_flux}
\end{figure*}

In addition, we analyze the NuSTAR observation, which was conducted entirely within the period covered by the XRISM observation studied here. We also analyze two NICER observations taken during the XRISM observation. The second was partially taken during the absorbed phase, and we studied only the portion before the end of the interval we chose. The observation log is given in Table \ref{log}, and the light curves and hardness ratios from NICER, NuSTAR, and XRISM Resolve and Xtend are shown in Figure \ref{lc}. The observations were taken during a moderately hard spectral state of Cyg X-1, as shown in Figure \ref{hr_flux}. This figure shows the count rates and hardness ratios from the MAXI X-ray monitor \citep{Matsuoka09}, with those corresponding to our chosen epoch marked in red. 

\begin{table*}
\caption{The log of the studied observations of Cyg~X-1 with NICER, NuSTAR, and XRISM}
  \centering
  \begin{tabular}{lcccccc}
   \hline
  Instrument & Obs ID & Start Time & Stop Time & Eff. Exp. \\
   & & (yyyy-mm-dd) & (yyyy-mm-dd) & (s)\\
   \hline    
     NICER 1 & 7100320101 & 2024-04-08 15:43:17 & 2024-04-08 21:59:48 & 871 \\
     NICER 2  &  7100320102 & 2024-04-09 01:02:36  &  2024-04-09 10:23:32 & 527 \\
      NuSTAR FPMA & 30901039002 & 2024-04-08 18:23:48 & 2024-04-09 06:26:06 & 16880 \\
        FPMB   & ...      & 2024-04-08 18:23:48 & 2024-04-09 06:26:06 & 17280 \\

     XRISM Resolve &  300049010 & 2024-04-07 17:56:58 &  2024-04-09 11:51:16 & 82490 \\
         Xtend & ... &  2024-04-07 17:43:02 & 2024-04-09 11:37:20 & 9440 \\
   \hline
  \end{tabular}
  \label{log}
\end{table*}

We processed the XRISM Resolve data using HeaSoft V6.36 and CALDB version 20250915, following the procedures described in \cite{Yamada25} and \dr. For spectral extraction with \texttt{xselect}, we considered only high-resolution primary (Hp) events and excluded pixels 12 and 27. The small (S) and extra-large (XL) response matrix files (RMFs) were generated with the \texttt{rslmkrmf} task after excluding low-resolution secondary (Ls) events. The ancillary response file (ARF) was then derived using the \texttt{xaarfgen} task, with the exposure map produced by \texttt{xaexpmap} as input. The light curves were obtained in the 2--5 keV, 5--10 keV, and 2--10 keV bands, considering only Hp events.

The Xtend data were processed over the same time interval as the Resolve, following the procedure outlined in \citet{Yamada25}. A circular region with a radius of 60 pixels was used to extract the source spectrum. The RMF was generated with the \texttt{xtdrmf} task, and the ARF was generated with \texttt{xaarfgen} after deriving the exposure map with \texttt{xaexpmap}. The relatively short exposure for Xtend was due to the screening criteria in the \texttt{WINDOW2BURST} mode. See also section 2.3 of \citet{Yamada25}.

The NuSTAR data were processed with the standard \texttt{nupipeline} task in NuSTARDAS, using the bright-source filtering option\footnote{\texttt{statusexpr=“(STATUS==b0000xxx00xxxx000)\&\&(SHIELD==0)"}}. Source and background spectra were extracted from circular regions of radius $100\arcsec$ with the \texttt{nuproducts} task, which also produced the corresponding RMF and ARF. The same task was used to generate light curves in the energy bands 3--5 keV, 5--10 keV, and 3--78 keV. 

The two NICER observations were processed with the standard \texttt{nicerl2} pipeline and with the {\tt night} setting. We then combined the first and second observations covering the interval of interest using the \texttt{niobsmerge}{\footnote{\url{https://heasarc.gsfc.nasa.gov/docs/nicer/analysis_threads/combine-obs/}}} task. This combined observation was used to extract the source and background spectra, along with the corresponding RMF and ARF, using the \texttt{nicerl3-spect} script. We adopted the ``3C50'' model for background estimation. During data reduction, the script added 1.5\% systematic error to each 0.34--9.0\,keV channel, and somewhat more (up to 2.5\%) beyond that\footnote{\url{https://heasarc.gsfc.nasa.gov/docs/nicer/analysis_threads/cal-recommend/}}. Light curves in the 0.3--2 keV, 2--10 keV, and 0.3--10 keV bands were extracted with the \texttt{nicerl3-lc} script. All spectral data were optimally binned \citep{Kaastra16}, with the additional requirement that each bin contain at least 25 counts.

We have calculated the fractional variability, rms($E$), as a function of photon energy over the frequency range 0.01--1 Hz using the \texttt{Stingray} package \citep{Huppenkothen2019a, Huppenkothen2019b}. For the NuSTAR data, we have used the co-spectrum computed with the \texttt{HENDRICS} package \citep{Bachetti15}. However, we have found that the NuSTAR data show significantly lower variability than NICER and Resolve in the overlapping 3--10 keV energy range. This appears to be due to dead time in the \nustar data, which artificially lowers the rms (even when the co-spectrum is used; \citealt{Bachetti18}; M. Bachetti, private communication). 

On the other hand, the NICER and Resolve rms values show only minor differences in the Fe K line energy band, likely arising from their different instrumental responses\footnote{Due to NICER's silicon drift detectors' relatively broad energy response, variable high-energy continuum photons can be redistributed into the Fe K energy range, thereby enhancing the apparent variability in this band. In contrast, XRISM's Resolve microcalorimeter, with its substantially higher spectral resolution, minimizes spectral mixing between the continuum and the narrow Fe K. As a result, the Fe K band in XRISM would show a lower fractional rms amplitude because it is less contaminated by variable continuum photons.}. Alternatively, the discrepancy may reflect intrinsic source non-stationarity combined with XRISM's longer temporal coverage. We also attempted to derive rms($E$) for the Xtend data. However, its GTI ('good time interval') structure was highly fragmented, and the GTI lengths were shorter than the segment lengths required to derive rms($E$). This was probably due to the \texttt{WINDOW2BURST} mode used.

\begin{figure}
\centerline{\includegraphics[width=0.9\columnwidth]{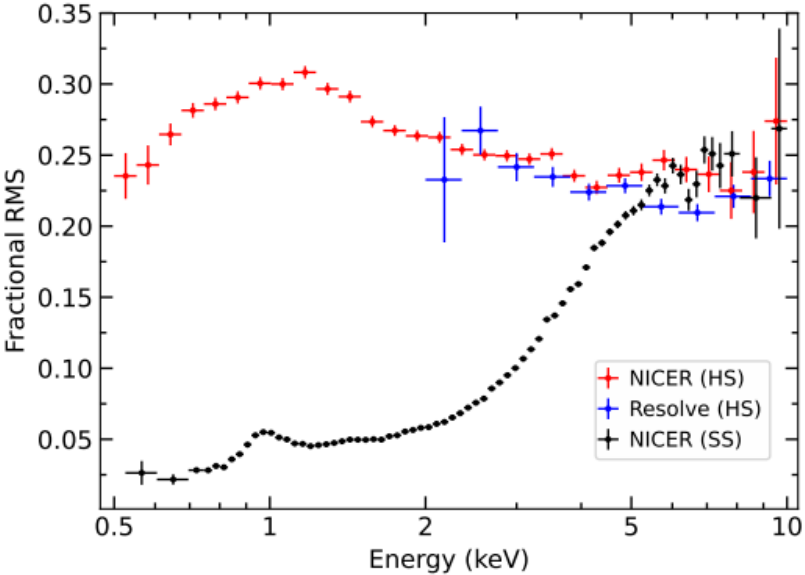}}
 \caption{The rms$(E)$ in the 0.01--1 Hz frequency range for NICER (red) and XRISM Resolve (blue). For comparison, the black symbols show the rms$(E)$ from NICER in the soft spectral state \citep{Zdziarski24b}. We see similar trends in soft X-rays, with broad peaks around 1 keV, though the soft-state rms is much lower. Above that, we observe a decrease with energy in the hard state and an increase in the soft state, while both converge in the Fe K energy region. 
}\label{rms}
\end{figure}

The results, excluding NuSTAR, are shown in Figure \ref{rms}. We see the rms increase up to about 1.2 keV, then decrease. Similar results for another NICER observation are shown by \citet{Basak25}. They calculate the rms separately for four frequency bands and show that the hump around 1.2 keV and the decline of the rms above it are most pronounced at frequencies $<$0.5 Hz. This behavior is consistent with the findings of \citet{WU09} for two other BH XRBs, which they interpreted as long-term variability of the coronal emission driven by disk variability (see also \citealt{Uttley11}). In that process, variable disk seed photons enter the corona, modulating its cooling and thus the spectral slope (see, e.g., figure 14 in \citealt{ZPPW02}). However, the decline of the rms at energies below $\sim$1 keV indicates the presence of a more stable disk component, which is difficult to explain in this scenario. For comparison, we show the rms spectrum of Cyg X-1 in the soft spectral state \citep{Zdziarski24b}. Curiously, there is a similar hump at $\approx$1 keV. Those authors proposed that it may be caused by photon transmission through the stellar wind.

\section{Spectral fitting}
\label{spectral}

Spectral fitting is performed using {\sc xspec} \citep{Arnaud96} with $\chi^2$ statistics. Uncertainties are calculated at the 90\% confidence level ($\Delta\chi^2 \approx 2.71$; \citealt{Lampton76}). We account for absorption using the {\tt tbfeo} model in {\sc xspec} and assume the absorber's elemental abundances from \citet{Wilms00}. In prior work, the column density toward Cyg X-1 was found to be $N_{\rm H}\approx (6\pm 2)\times 10^{21}$ cm$^{-2}$ based on reddening toward the donor \citep{Balucinska95}, and $\approx$4.65--$4.85\times 10^{21}$ cm$^{-2}$ based on modeling of dust scattering toward Cyg X-1 \citep{Xiang11}. \citet{Tomsick14} fitted the soft state of Cyg X-1, including absorption by the stellar wind. That state has strong soft emission, allowing a relatively precise determination of the column. They found $N_{\rm H}\approx 6.6\pm 0.2$ and $6.5\pm 0.1\times 10^{21}$ cm$^{-2}$ for their two models with a relativistic disk. Then, \citet{HI4PI} obtained $N_{\rm H}\approx 7\times 10^{21}$ cm$^{-2}$. Based on those works, we decided to constrain $N_{\rm H}\geq 5\times 10^{21}$ cm$^{-2}$.  

When fitting multiple data sets, we account for residual spectral calibration differences using the {\tt plabs} model in {\sc xspec}. It multiplies the model spectra by $K E^{-\Delta\Gamma}$, with $K=1$ and $\Delta\Gamma=0$ for the \nustar A and B units, and we allow $K$ and $\Delta\Gamma$ to vary freely for the NICER and XRISM Resolve data sets. We note that {\tt plabs} is equivalent to the {\tt CRABCORR} model introduced by \citet{Steiner10} for the same purpose. 

\subsection{XRISM Resolve data}
\label{Resolve}

We have found that our Resolve spectrum lies significantly above the NICER data below 2.6 keV (see Appendix \ref{cross}). We therefore chose the energy range 2.6--12 keV for spectral fitting. By contrast, \dr used the 2--12 keV range because their data showed no systematic residuals below 2.6 keV. This minor difference may be due to the choice of the observation interval or details of the data extraction procedure. Given the fitted energy range, we can only weakly constrain X-ray absorption, so we use the {\tt tbfeo} absorber model with the O and Fe relative abundances fixed at unity. 

Since our aim is to constrain the continuum and the broad Fe K spectral components, we used the small (S) spectral response files. We used the results of \dr to model narrow emission lines present in the spectrum (see their figure 3). The figure shows two slightly broadened Fe K lines at 6.39 and 6.40 keV. We fitted them with Gaussians and, in subsequent analyses, kept their energy and width fixed at the best-fit values. We verified that their parameters have virtually no effect on the continuum-fitting results. In our initial analyses, we assumed the disk extends down to the ISCO and fit the spin parameter. 

\setlength{\tabcolsep}{2pt}
\begin{table}
\caption{The results of spectral fitting of the Resolve data alone}
\vskip -0.4cm
   \centering\begin{tabular}{lcccc}
\hline
Parameter & Symbol & {\tt relxill} & {\tt relxillCp} & {\tt reflkerrD} \\
&[Unit]&& &\\
\hline
ISM & $N_{\rm H}$ $[10^{21}$\,cm$^{-2}$] & $5.8^{+3.2}_{-0.8}$ & $5.0^{+0.7}$ & $7.6_{-2.1}^{+3.1}$  \\ 
Spin & $a_*$ & $0.988^{+0.003}_{-0.008}$ & $0.00^{+0.17}$ & $0.18_{-0.18}^{+0.10}$\\
Inclination & $i\,[\degr]$ & $70^{+3}_{-5}$ & $33^{+1}_{-2}$ & $36^{+1}_{-4}$ \\
Irradiation  & $q_1$ & $8.5^{+1.0}_{-0.3}$ & $10.0_{-5.9}$ & $10.0_{-4.7}$ \\
\, \, Indices & $q_2$  & $0.0^{+1.0}_{-1.1}$ &  $2.9^{+0.4}_{-0.4}$     & $2.5^{+0.4}_{-0.3}$\\
Break radius & $R_{\rm br}\,[R_{\rm g}]$ & $7.8^{+9.0}_{-3.4}$ & $7.6^{+3.1}_{-1.3}$ & $7.6^{+1.7}_{-2.4}$  \\
Power law & $\Gamma$  & $1.76^{+0.09}_{-0.12}$ & $1.73^{+0.01}_{-0.02}$\\
Compton & $y$ &&&$1.17^{+0.07}_{-0.05}$\\
Cutoff & $E_{\rm c}$\,[keV] &  $120^{+200}_{-60}$\\
Temperature & $kT_{\rm e}$\,[keV] &  & $400_{-320}$ & $84^{+7}_{-9}$  \\
Blackbody & $kT_{\rm bb}$\,[keV] & -- & 0.05f & $0.43_{-0.03}^{+0.03}$\\ 
Reflection & ${\cal R}$ & $2.0_{-0.4}$ & $0.4^{+0.1}_{-0.1}$ &$0.28^{+0.11}_{-0.01}$\\
Density & $\log_{10} n$\,[cm$^{-3}$] & -- & $20.0_{-0.4}$ & $15.0^{+4.0}$\\
Ionization   &$\log_{10}\xi$\,[erg\,cm\,s$^{-1}$] & $2.8^{+0.1}_{-0.5}$ & $2.8^{+0.1}_{-0.2}$ &    $3.2^{+0.1}_{-0.1}$  \\
Abundance & $Z_{\rm Fe}\,[\sun]$ & $0.7^{+0.2}_{-0.2}$ &  $1.4_{-0.3}^{+0.5}$ & $3.0_{-1.0}$\\
\hline
Fit & $\chi_\nu^2$  & 3479/3365 & 3481/3364 & 3473/3363\\
\hline
\end{tabular}
\tablecomments{We constrained $0\leq a_*\leq 0.998$, $N_{\rm H}/10^{21}{\rm cm}^{-2}\geq 5$, $0\leq q_{1,2}\leq 10$, $15\leq \log_{10} n\leq 20$ in {\tt relxillCp} and $15\leq \log_{10} n\leq 19$ in {\tt reflkerrD}, $\log_{10} \xi\leq 4.7$, $kT_{\rm e}\leq 400$\,keV, $Z_{\rm Fe}\leq 3$. The ionization parameter, $\xi$, is defined in Equation (\ref{xi}). The Comptonization $y$ parameter is defined as $y\equiv 4 kT_{\rm e}\tau$, where $\tau$ is the Thomson optical depth of the corona. This parameter approximately determines the slope of the power-law portion of the Comptonization spectrum. The reflection fraction, ${\cal R}\leq 2$, is defined as the ratio of the flux incident on the disk to that escaping to infinity. 
}
\label{Resolve_fits}
\end{table}

We first used the {\tt relxill} model \citep{GK10, Dauser10, Garcia13}, which calculates the reflection of an e-folded power-law spectrum from a disk with a density of $n=10^{15}$ cm$^{-3}$ (typical for disks in active galactic nuclei) and a broken-power-law dependence of the irradiating flux, $\propto R^{-q_{1,2}}$ below and above a break radius, $R_{\rm br}$. While \dr used {\tt relxill} v.2.3, we use the current version, v.\ 2.8. We obtained results very similar to those of \dr (see Table \ref{Resolve_fits}). In particular, we obtain the spin very close to the maximum, $a_*=0.988^{+0.003}_{-0.008}$, confirming the finding of \dr. We also obtain a similarly high disk inclination, $i=70^{+3}_{-5}\degr$ (while \dr obtained $i=62^{+1}_{-1}\degr$). Such high values were rarely obtained earlier; e.g., the spectral fitting of \citet{Tomsick18} gave $i\approx 35$--$38\degr$, while \citet{Zdziarski24b} found $i=39^{+1}_{-1}\degr$. Also, the current values of the orbital inclination are in the 27--$34\degr$ range (Section \ref{intro}). This model also yields a rather strong reflection. We constrained the relative reflection strength to ${\cal R}\leq 2$ (to allow for a moderate enhancement due to Compton anisotropy; \citealt{Ghisellini91}), and obtained ${\cal R}= 2.0_{-0.4}$. \dr obtained a similarly high value of ${\cal R}= 1.7\pm 0.1$. 

\begin{figure}
\centerline{\includegraphics[width=0.8\columnwidth]{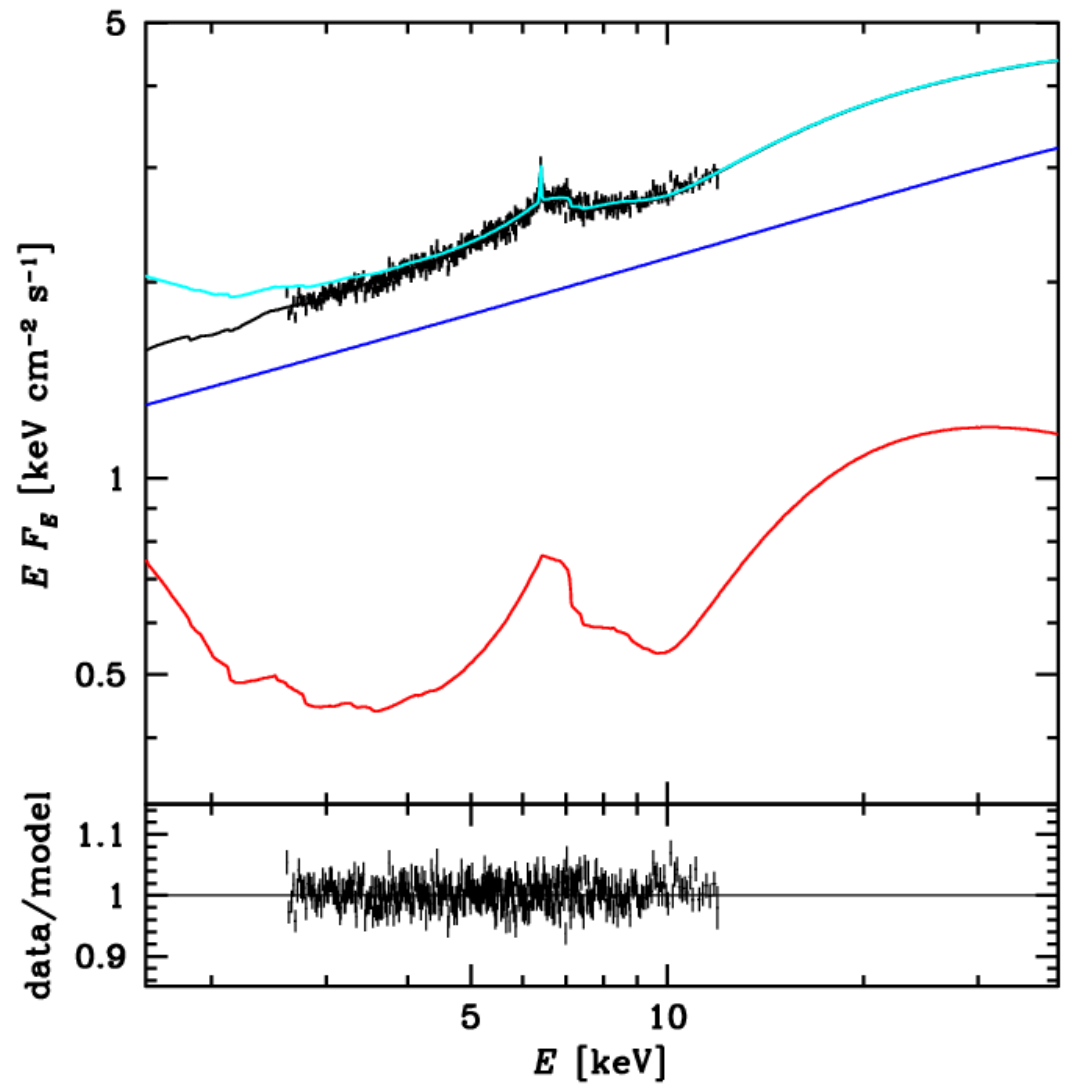}}
\caption{The Resolve unfolded spectrum (top) and data-to-model ratio (bottom) for the fit using {\tt relxillCp}. The total and unabsorbed model spectra are shown by the solid black and cyan curves, respectively. The incident and reflection components are shown by the blue and red curves, respectively. Here and in Figures \ref{2reflkerrD} and \ref{full} below, the spectra have been rebinned for plotting. 
}\label{relxillCp}
\end{figure}

We then use the {\tt relxillCp} model \citep{Garcia18}, which uses a thermal Comptonization spectrum from \citet{ZJM96} as the incident spectrum and allows variable disk density. This is a more advanced and physically motivated model than {\tt relxill}. It is better suited to Cyg X-1 because its hard-state spectra at $\lesssim$100 keV are better modeled by thermal Comptonization (e.g., \citealt{Gierlinski97, McConnell02}) than by the e-folded power law assumed in {\tt relxill}. It assumes the temperature of blackbody seed photons undergoing Comptonization is 50 eV  \citep{Garcia18}, as shown in Table \ref{Resolve_fits}. While that temperature in Cyg X-1 appears to be higher (see Section \ref{broad} below), {\tt relxillCp} can be used at energies much above those of the seed photons, which is the case for the Resolve data. 

We find results that differ significantly from those of {\tt relxill} and present them in Table \ref{Resolve_fits} and Figure \ref{relxillCp}. We first allow negative spin values and obtain a slightly negative best-fit spin, $a_*=-0.09^{+0.22}_{-0.11}$. Since the donor is close to filling its Roche lobe (e.g., \citealt{Ramachandran25}), retrograde accretion is unlikely, so we constrain $a\geq 0$, which increases $\chi^2$ by less than 1. With this constraint, the spin is $a_*=0.00^{+0.17}$, compatible with the gravitational-wave results. The inclination is $i=33^{+1}_{-2}\degr$, consistent with the measured orbital inclination. The reflection is weak, ${\cal R}= 0.4\pm 0.1$, consistent with an outflowing corona \citep{Poutanen23} that can explain the X-ray polarization measured in this source \citep{Krawczynski22}. The Fe abundance is now $>1$, consistent with an advanced evolutionary stage of the donor of Cyg X-1. The disk density is high, $\geq 10^{20}$ cm$^{-3}$, in agreement with the result for Cyg X-1 from \citet{Tomsick18}. The $\chi^2_\nu$ is only slightly larger, by +2, than that of the previous model. We have also checked the effect of including energies down to 2 keV in the fit. This produced only systematic, mostly positive residuals below 2.6 keV, with virtually no effect on the results. In particular, $a_*=0.00^{+0.18}$ was found. 

We also consider another advanced reflection model, {\tt reflkerrD} \citep{Niedzwiecki19}. It uses the Comptonization model of \citet{PS96} for the incident spectra, which is significantly more accurate than the model of \citet{ZJM96} used by {\tt relxillCp}. It also allows a free temperature for the seed photons, which provides greater self-consistency than the $kT_{\rm bb}=50$ eV assumed by {\tt relxillCp}. However, {\tt reflkerrD} allows $a\geq 0$ only. We obtain spin and inclination values very similar to those from {\tt relxillCp} (see Table \ref{Resolve_fits}) and a $\chi^2_\nu$ lower than that of the previous models. However, the Fe abundance is relatively high, $\sim$3. This appears to be related to an older version of the reflection tables (see \citealt{Garcia18n}) used by {\tt reflkerrD} compared to those used by the current version of {\tt relxillCp}.

We then searched for alternative solutions using {\tt relxill} at low spin and {\tt relxillCp} at high spin. For the former, there is a solution with $a_*=0$ (and at $i\approx 35\degr$) but with a significantly higher $\chi^2=3497/3365$. For the latter, there is a solution with $a_*\approx 0.9$ and $\chi^2=3489/3364$. Thus, these solutions provide significantly worse fits than the main, global solutions.

We then tested disk truncation. For {\tt relxill}, allowing the inner radius to vary does not improve the fit; $a_*$ remains nearly unchanged. The disk inner radius is constrained to 1.0--$1.26 R_{\rm ISCO}$, with the upper limit corresponding to the maximum allowed spin, $a_*=0.998$ \citep{Thorne74}, at which $R_{\rm ISCO}=1.237 R_{\rm g}$ \citep{Bardeen72}. Equivalently, $R_{\rm in}\lesssim 1.6 R_{\rm g}$. 

Although allowing the inner radius to be free does not improve the fit with {\tt relxillCp}, the resulting constraints on $a_*$ and $R_{\rm in}$ differ. We find that the spin parameter is then completely unconstrained, i.e., it can range from $a_*=0$ to 1, whereas the allowed range of the truncation radius is limited to a narrow interval in $R_{\rm g}$. For a fixed $a_*=0$ (with $R_{\rm ISCO}=6 R_{\rm g}$), $R_{\rm in}=6.2_{-0.2}^{+0.3}R_{\rm g}$, whereas for $a_*=0.99$ (with $R_{\rm ISCO}\approx 1.45 R_{\rm g}$), $R_{\rm in}=5.5_{-0.5}^{+0.3}R_{\rm g}$, which yields an unchanged $\chi^2$ in both cases. We also obtained very similar ranges with {\tt reflkerrD}. This has a remarkable implication. Based on the Resolve data alone, the disk's inner radius is limited to the ISCO radius of a slowly spinning BH. Thus, the spin is unconstrained; either the BH is slow-spinning, and the disk extends down to the ISCO, or there is a modest disk truncation for a fast-spinning BH. 

\subsection{Broadband spectra}
\label{broad}

\begin{table}
\caption{The results of spectral fitting of the joint NICER, Resolve, and NuSTAR data}
   \centering\begin{tabular}{lccc}
\hline
Parameter & Symbol & {\tt reflkerrD} & 2$\times${\tt reflkerrD} \\
&[Unit]&& + {\tt diskbb}\\
\hline
ISM & $N_{\rm H}$ $[10^{21}$\,cm$^{-2}$] & $5.8^{+0.2}_{-0.1}$ & $7.6^{+1.7}_{-0.5}$  \\ 
& $Z_{\rm O}\,[\sun]$ & $1.22^{+0.07}_{-0.03}$ & $1.30^{+0.05}_{-0.07}$\\
& $Z_{\rm Fe}\,[\sun]$ & $0.50^{+0.02}$ & $0.88_{-0.16}^{+0.14}$ \\
Spin & $a_*$ & $0.10^{+0.90}_{-0.10}$ & $0.18^{+0.82}_{-0.18}$\\
Inclination & $i\,[\degr]$ & $41^{+1}_{-1}$ & $32^{+1}_{-1}$ \\
Irradiation  & $q_1$ & $0.0^{+0.4}$ & $0.0^{+0.8}$ \\
Temperature & $kT_{\rm e}$\,[keV] &  $110_{-2}^{+12}$ & $34^{+1}_{-1}$  \\
Blackbody & $kT_{\rm bb}$\,[keV] & $0.30_{-0.01}^{+0.01}$ & $0.15_{-0.01}^{+0.01}$\\ 
Disk & $N\,[10^5]$ & -- & $9.3^{+1.1}_{-3.1}$\\
Density & $\log_{10} n$ & $16.9_{-0.4}^{+0.2}$ & $18.2^{+0.4}_{-0.1}$\\
Abundance & $Z_{\rm Fe}\,[\sun]$ & $3.0_{-0.2}$ & $2.8^{+0.2}_{-0.1}$\\
Compton & $y$ & $1.28^{+0.01}_{-0.01}$ & $2.17^{+0.03}_{-0.04}$\\
Ionization   &$\log_{10}\xi$ & $2.8^{+0.1}_{-0.1}$ &    $4.3^{+0.1}_{-0.1}$  \\
Reflection & ${\cal R}$ & $0.21^{+0.01}_{-0.01}$ &$0.48^{+0.05}_{-0.09}$\\
Power law 2 & $y_2$ & -- & $0.71^{+0.05}_{-0.02}$\\
Ionization 2   &$\log_{10}\xi_2$ & -- & $3.3^{+0.1}_{-0.1}$  \\
Reflection 2 & ${\cal R}_2$ & -- &$0.37^{+0.01}_{-0.04}$\\
Cross-&$\Delta\Gamma_{\rm NICER}$&$-0.01^{+0.01}_{-0.01}$& $-0.11^{+0.01}_{-0.01}$\\
calibration& $K_{\rm NICER}$ & $0.96^{+0.01}_{-0.02}$ & $0.76^{+0.02}_{-0.01}$\\
&$\Delta\Gamma_{\rm Resolve}$&$0.09^{+0.01}_{-0.01}$& $0.00^{+0.02}_{-0.02}$  \\
& $K_{\rm Resolve}$ & $1.09^{+0.02}_{-0.03}$ & $0.88^{+0.01}_{-0.01}$\\
\hline
Fit & $\chi_\nu^2$  & 4247/3975 & 4106/3970\\
\hline
\end{tabular}
\tablecomments{The constraints are the same as in Table \ref{Resolve_fits}.}
\label{fits_broad}
\end{table}

We now perform spectral fitting of the joint spectrum from NICER, NuSTAR, and XRISM Resolve over the 0.5--79 keV range. We have found that the Resolve spectrum contains a significantly stronger Fe K complex than either the NuSTAR A or B spectra, while the NICER spectrum in that range agrees with the Resolve spectrum within the uncertainties (see Appendix \ref{cross}). Therefore, we used only the NuSTAR spectra in the 9--79 keV range (similar to \dr). 

\begin{figure}
\centerline{\includegraphics[width=0.8\columnwidth]{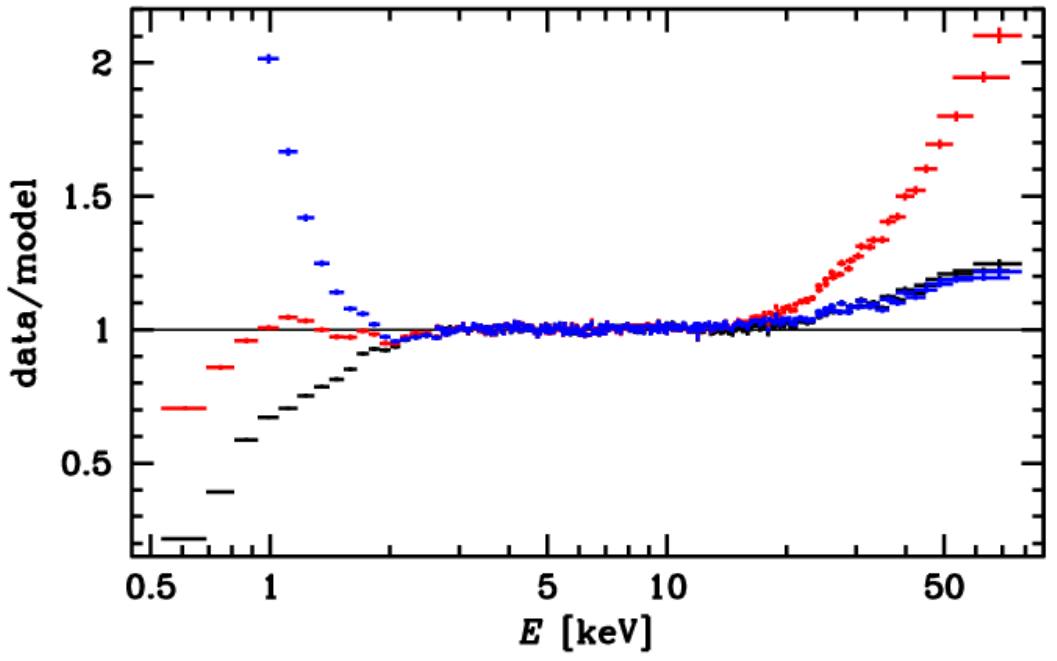}}
\caption{The data-to-model ratios for the models fitted to the Resolve data alone are shown for the broadband data. The red, blue, and black symbols correspond to the {\tt relxill}, {\tt relxillCp}, and {\tt reflkerrD} models, respectively. The NICER and NuSTAR data are renormalized using the cross-calibration coefficients of our best broadband model (see the last column in Table \ref{fits_broad}). We see that all of the models fitted to 2.6--12 keV Resolve data diverge from the data beyond that range.
}\label{ratios}
\end{figure}

Figure \ref{ratios} demonstrates the strong need for broadband data. It shows the data-to-model ratios for the three models fitted to the Resolve data alone across the full available energy range. We see that while all the models fit well in the 2.6--12 keV range, they diverge strongly from the data beyond that range.

In our approach, we treat blackbody photons emitted by the accretion disk as seeds for Comptonization in the corona. This choice limits our model options. In particular, the seed photon temperature of {\tt relxillCp} is fixed at 50 eV (see Section \ref{Resolve}), which is not applicable to our spectra. We therefore use only the {\tt reflkerrD} model, which allows the seed blackbody temperature to vary freely. We also use its option in which the seed photons have a disk blackbody spectrum (with a maximum temperature of $T_{\rm bb}$) from the {\tt diskbb} model \citep{Mitsuda84}.

\begin{figure}
\centerline{\includegraphics[width=\columnwidth]{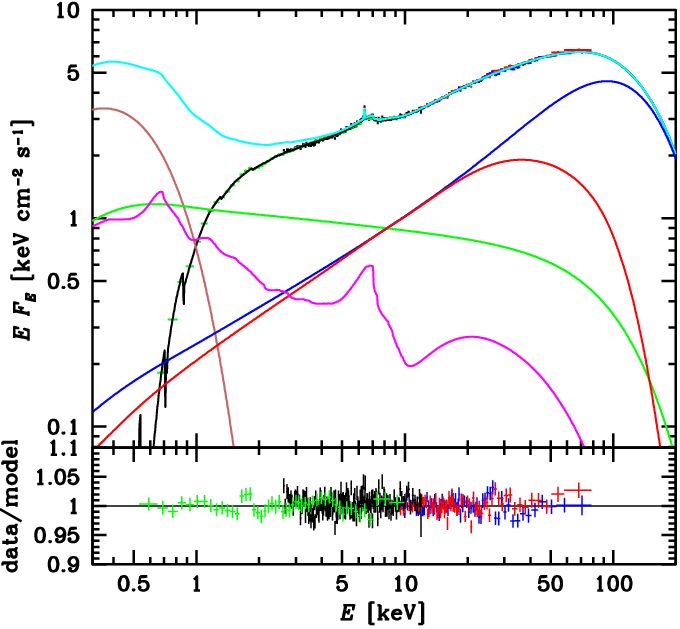}}
\caption{The NICER (green), Resolve (black), and NuSTAR (blue and red) unfolded spectra (top) and data-to-model ratios (bottom) for the fit with two {\tt reflkerrD} components. The total and unabsorbed model spectra are shown by the solid black and cyan curves, respectively. The incident and reflection components are shown by the blue and red curves for the hard component, and by the green and magenta curves for the soft component. The unscattered disk component is shown by the brown curve.
}\label{2reflkerrD}
\end{figure}

The soft X-rays emitted by Cyg X-1 are absorbed by both the ISM and the donor's stellar wind. The latter was taken into account by \citet{Tomsick14} using a table model based on the {\sc xstar} ionization model \citet{xstar}. We tested that model. However, we found that including it has virtually no effect on the derived parameters, and the fit converges at the lowest column density allowed by this model. Thus, we decided to model absorption only with {\tt tbfeo}. However, we allow the O and Fe abundances in that model to differ from unity. We also found that allowing $q_2$ and $R_{\rm br}$ to be free has little effect on the final fits, so we kept them at the default values, $q_2=3$ and $R_{\rm br}=15 R_{\rm g}$. 

We first consider a model with a single Comptonization component, as in the fits to the Resolve data alone. We obtained an unconstrained spin of $a_*=0.10^{+0.90}_{-0.10}$ and an inclination of $i=42^{+1}_{-1}\degr$ at $\chi^2_\nu=4247/3975$ (see Table \ref{fits_broad}). However, fitting the broadband spectrum prevents this model from properly accounting for the Resolve spectrum. The $\chi^2$ contribution to the broadband spectrum is 3581, which is by $>$100 higher than the $\chi^2$ of the models used for the Resolve spectrum alone (see Table \ref{Resolve_fits}). The fit also shows substantial wavy residuals (especially below 3 keV) and fails to reproduce the Fe K complex. We have also added a separate {\tt diskbb} component (with the inner temperature equal to that used for Comptonization) to represent the unscattered disk emission. However, it did not improve the fit, and the best-fit {\tt diskbb} normalization was null. 

We have therefore added a second Comptonization component, first to the model without {\tt diskbb}. We assumed that the electron and disk temperatures, the density, and the Fe abundance were the same as in the first component. This adds four free parameters. The fit shows a highly significant improvement, with $\chi^2_\nu= 4153/3971$. Furthermore, we added a separate {\tt diskbb} component in the same way as for the model with single Comptonization. In contrast to that model, this further improved the fit by $\Delta \chi^2 =- 47$ (with the addition of one free parameter). The final $\chi^2_\nu=4106/3970$; see Table \ref{fits_broad}. The formal F-test probability that the improvement with respect to the single-Compton model is by chance is $\approx 1\times 10^{-27}$. However, as discussed by \citet{Protassov02}, while the large decrease in $\chi^2$ indicates a much improved fit, that probability should not be interpreted as the actual one. The $\chi^2$ contribution to the Resolve data is 3495, with a moderate increase relative to the models for those data alone. The resulting parameters are now similar to those obtained for the Resolve data alone using either {\tt relxillCp} or {\tt reflkerrD}, except that the spin parameter is now unconstrained, $a_*=0.19^{+0.81}_{-0.19}$. Figure \ref{2reflkerrD} shows the unfolded spectrum, the residuals, and the fit components. We find that the Fe K complex is dominated by reflection from the softer Comptonization component. The ionization parameter of the reflecting medium for the harder Comptonization is very high, and its reflection spectrum is virtually featureless. We note that the model's maximum disk temperature, $0.15_{-0.01}^{+0.11}$ keV, fully agrees with that obtained by \citet{Basak17}.

We note that the $\Delta\Gamma$ coefficients differ between models with one and two Comptonization components, inconsistent with their interpretation as cross-calibration coefficients. The values for the former model appear to reflect its shortcomings, as discussed above. For the latter, a negative $\Delta\Gamma_{\rm NICER}$ is consistent with findings from another joint observation of Cyg X-1 by NICER and NuSTAR \citep{Zdziarski24b}. On the other hand, the Resolve and NuSTAR spectral slope calibrations agree fully for the latter: $\Delta\Gamma_{\rm Resolve}=0.00^{+0.02}_{-0.02}$. The values of $K$ reflect both possible differences in flux normalization and source variability, given the instruments' different exposure times; see Table \ref{log}.

Next, we test for disk truncation. We find results similar to those obtained with {\tt relxillCp} for the Resolve data alone. Allowing truncation reduces the $\chi^2$ only slightly and is not statistically required. Since the spin is unconstrained in this model, we consider two values: $ a_*=0.1$ and $0.998$. We then find that the disk inner radius must be $R_{\rm in}\lesssim 10 R_{\rm g}$ for the former and $\lesssim 8 R_{\rm g}$ for the latter. We also find that the disk inner radius is limited to $\lesssim 10 R_{\rm g}$ at any spin. 

However, we find that the irradiation index for the model with the disk extending down to the ISCO is very low, $ q_1 \approx 0.0^{+0.8}$. This indicates that the innermost regions around the BH contribute very little to the observed emission. We have therefore tested a modified model with a constant $q_1=3$. In that case, the best fit still shows truncation, with $R_{\rm in}\approx 11 R_{\rm g}$, $a_*\approx 0.76$, and $\Delta\chi^2\approx +2$ relative to the model with truncation and a free $q_1$. Thus, a moderate disk truncation together with the standard irradiation remains a viable possibility.

\section{Disc-corona Energetics}
\label{energetics}

The unabsorbed bolometric flux of our two-Compton model for the broadband spectrum is $\approx 5.2\times 10^{-8}$ erg cm$^{-2}$ s$^{-1}$. This is typical of the hard state (see Table 1 in \citealt{ZPPW02}). Assuming isotropy and $D=2.2$ kpc, this corresponds to $\approx 3.0\times 10^{37}$ erg s$^{-1}$. The Eddington luminosity for an H abundance of 0.5 (characteristic of its donor star) is (2.3--$3.5)\times 10^{39}$ erg s$^{-1}$ for a BH mass in the range of 14--$21\msun$ (see Section \ref{intro}). Thus, the Eddington ratio is $\sim$1\%. 

The normalization of the {\tt diskbb} component for this model is given in Table \ref{fits_broad}. However, this represents only the unscattered disk flux. We have calculated the normalization of the {\tt diskbb} emission that is scattered by our two Compton clouds by fitting the incident model spectra with {\tt compps}, the Comptonization model assumed in {\tt reflkerrD}. The total normalization is then $N_{\rm t}\approx 1.2\times 10^6$. Using the definition in {\tt diskbb}, this can be converted to the disk inner radius by
\begin{equation}
R_{\rm in}=10^5 f_{\rm in} f_{\rm col}^2 \frac{D}{10\,{\rm kpc}} \left(\frac{N_{\rm t}}{\cos i}\right)^{1/2}\,{\rm cm},
\label{rin_disk}
\end{equation}
where $f_{\rm col}\approx 1.3$--1.9 is the color correction \citep{Davis19}, and $f_{\rm in}<1$ is a correction factor that accounts for {\tt diskbb} omitting the zero-stress term at the ISCO (for cases where the disk extends to the ISCO). For our values of $N_{\rm t}$, $D$ and $i$, this corresponds to 
\begin{equation}
\frac{R_{\rm in}}{R_{\rm g}}\approx 6 \left(\frac{M}{21\msun}\right)^{-1} \frac{f_{\rm in}}{0.41} \left(\frac{f_{\rm col}}{1.3}\right)^2,
\label{rin_solution}
\end{equation}
where we scaled it to the value of $f_{\rm col}$ for the fitted $kT_{\rm bb}$ and an Eddington ratio of 0.01 (see figure 5 of \citealt{Davis19}), and to $f_{\rm in}=0.41$ obtained by \citet{Kubota98}. For $M=14\msun$, the coefficient above is 9. Equation (\ref{rin_solution}) indicates either the inner radius at the ISCO of a slowly rotating BH or a slightly truncated disk. Importantly, it is inconsistent with a disk extending to the ISCO of a maximally rotating BH. A caveat is that {\tt diskbb} assumes a constant irradiation index of 3, whereas we allowed it to be free in an inner region. Also, the value of $f_{\rm in}=0.41$ used above was calculated only for $a_*=0$ \citep{Kubota98}, and $f_{\rm in}\sim 1$ if the disk is truncated above the ISCO. Also, $f_{\rm col}$ can be $>1.3$. Taking that into account, we have a constraint of $R_{\rm in}>(6$--$9)R_{\rm g}$ for the considered range of $M$. 

The disk flux is
\begin{equation}
    F_{\rm disk}=2\left(10^4\,{\rm cm}\over 1\,{\rm kpc}\right)^2 N_{\rm t}\sigma T_{\rm bb}^4\approx 1.3\times 10^{-8}\,{\rm erg}\,{\rm cm}^{-2}\,{\rm s}^{-1},
\end{equation}
where $\sigma$ is the Stefan-Boltzmann constant. This can be compared with the observed flux in the coronal Comptonization spectra, $F_{\rm cor}\approx 2.5\times 10^{-8}$ erg cm$^{-2}$ s$^{-1}$. Taking 0.4 as the average observed reflection fraction and assuming an average albedo of 0.5 (given that the reflection of the dominant hard component arises from a strongly ionized medium), the irradiation-deposited flux is $0.4(1-0.5) 2.5\times 10^{-8}$ erg cm$^{-2}$ s$^{-1}\approx 5\times 10^{-9}$ erg cm$^{-2}$ s$^{-1}$. Thus, the disk emission has comparable contributions from both intrinsic dissipation and irradiation. On the other hand, the coronal flux, $F_{\rm cor}$, is about twice that of the disk, $F_{\rm disk}$. Also, as shown by comparing the total {\tt diskbb} normalization, $N_{\rm t}$, to that of the unscattered disk emission, $N$, the corona covers a relatively small fraction of the disk, $(N_{\rm t}-N)/N_{\rm t}\approx 0.3$.

We can also check the self-consistency among the flux impinging on the disk, the disk density, and the ionization parameter, following \citet{ZDM20}. We consider only the reflected flux of the softer Comptonization component in the fit to the broadband data, which equals $F_{\rm refl,2}\approx 6\times 10^{-9}$ erg cm$^{-2}$ s$^{-1}$, and the corresponding isotropic luminosity is $4\pi D^2 F_{\rm refl,2}$. This flux is responsible for most of the observed Fe K component and thus must originate from the vicinity of the BH. Assuming it irradiates an area of $\sim 4\pi 10 R_{\rm g}^2$ (on both sides of the disk), we obtain the flux on the disk surface as $F_{\rm irr}=F_{\rm refl}D^2/(10 R_{\rm g})^2$. The ionization parameter is defined as
\begin{equation}
    \xi\equiv 4\pi F_{\rm irr}/n.
    \label{xi}
\end{equation}
This yields $\log_{10}\xi\approx 3.4$--3.7 for $M=21\msun$ to $14\msun$. Given the simplified and approximate nature of this estimate, it is in good agreement with the value of $3.3\pm 0.1$ in Table \ref{fits_broad}. 

\section{Discussion and Summary}
\label{discussion}

Our analysis has significantly extended the important work of \dr. While their study covered many details of the Fe K emission from Cyg X-1, particularly the narrow Fe K$\alpha$ features measured for the first time by XRISM Resolve, it used only a single model for the relativistically broadened X-ray reflection spectrum, {\tt relxill} \citep{Dauser10, Garcia13}. This deviated from the previous practice of those authors \citep{Draghis23, Draghis24, Draghis25a}, who considered and tested a range of available relativistic reflection models, particularly the improved, physically motivated model {\tt relxillCp} \citep{Garcia18}.

In our study, we have found a strong dependence of the fitted spin parameter on the model. While the {\tt relxill} model \citep{Garcia13} fitted to the Resolve data alone yields a spin parameter close to the maximal value, {\tt relxillCp} yields $a_*=0^{+0.17}$. Similarly, fitting with another physically motivated reflection model, {\tt reflkerrD} \citep{Niedzwiecki19}, yields $a_*=0.18_{-0.18}^{+0.10}$. An argument for the low spin is that the high disk inclination obtained with the {\tt relxill}, $\gtrsim\! 60\degr$, is inconsistent with both the binary inclination and most previous spectral-fitting results for Cyg X-1. By contrast, fitting with {\tt relxillCp} and {\tt reflkerrD} yields $i=33_{-2}^{+1}\degr$ and $i=36_{-4}^{+1}\degr$, respectively, i.e., inclination ranges compatible with the binary inclination of \citet{Ramachandran25}. As a possible explanation of the high disk inclination, a precession of the BH spin axis was invoked by \dr. Such precession would then govern the disk inclination and, in principle, could alter it relative to that reported in previous studies. However, the actual period of that de Sitter, spin-orbit, precession is several thousand years, see equation (14) of \citet{Zdziarski23b}. 

We then studied the broadband spectra from NICER (0.5--10 keV), XRISM Resolve (2.6--12 keV), and NuSTAR (9--79 keV), with the Resolve spectrum chosen to dominate the Fe K range. We found that models consisting of a single plasma cloud that Compton-upscatters photons from the disk and reflection cannot properly account for the observed spectrum. Instead, models including two Comptonizing regions, their reflections, and an accretion disk provide a relatively good description of the broadband data. The geometry can then be similar to that found to describe the hard state of MAXI J1820+070 (see figure 4 of \citealt{Zdziarski21b}). The harder component is closer to the BH, as evidenced by hard X-ray lags \citep{Kotov01, Basak25}, but it can also illuminate outer disk regions. 

However, the spin parameter in this model is unconstrained, $a_*=0.18_{-0.18}^{+0.82}$. By contrast, including the full energy range of the NuSTAR detector (3--79 keV) in the broadband fit (as described in Appendix \ref{cross}) yields a low spin, $a_*=0.10_{-0.10}^{+0.19}$. Thus, the fitted spin depends on both the model and the data set. Nevertheless, our results are consistent with low spin and, therefore, with the spin measurements from gravitational waves (see Section \ref{intro}). 

A notable result from our reflection spectroscopy is that the disk in Cyg X-1 appears at most weakly truncated, with the inner disk radius $R_{\rm in}\lesssim 10 R_{\rm g}$. Thus, the disk either extends to the ISCO of a slowly rotating BH or is weakly truncated. By contrast, the normalization of the disk component in our broadband modeling implies that its inner radius is several $R_{\rm g}$ {\it or more}. Taken together, these results indicate that the disk extends down to $\sim 10 R_{\rm g}$, consistent with the ISCO of a slowly rotating BH but not with that of a fast-rotating one. Thus, the data primarily constrain the truncation radius rather than the spin, and imply that if $a_*\sim 1$ in Cyg X-1, its disk is truncated. The disk's proximity to the BH is also consistent with the weak soft X-ray lags in Cyg X-1, as reported by \citet{Basak25}. The amplitude of the lag is $\approx$1 ms (see figure 10 in \citealt{Basak25}). If due to reverberation, this corresponds, at face value, to the light travel time over $\approx$(10--$15) R_{\rm g}$, depending on the BH mass.  

At the same time, the corona in the hard state of Cyg X-1 extends horizontally, as indicated by the X-ray polarization angle aligned with the jet of this source \citep{Krawczynski22}. Thus, the source geometry is that of a slab corona. However, static coronae exhibit soft, $\Gamma>2$, X-ray spectra due to feedback between the disk and corona \citep{HM91, PVZ18}, whereas we observe a hard spectrum with $\Gamma\sim 1.7$ (see Table \ref{Resolve_fits}). This discrepancy can be resolved if the corona is outflowing at a mildly relativistic velocity \citep{Beloborodov99}. Our results are consistent with the outflowing corona model of \citet{Poutanen23} in the variant where the underlying disk extends near the ISCO. The reflection fractions in our models are significantly less than unity due to relativistic beaming away from the disk. Our finding that the inner parts of the disk are only weakly irradiated, $\propto R^{-q_1}$ with $q_1\approx 0^{+0.8}$ (see Table \ref{fits_broad}), is consistent with a mildly relativistic outflow of the corona, which then reduces the irradiation of regions below the corona's edges. In addition, the model of \citet{Poutanen23} explains the high degree of linear polarization in X-rays, $4.0\%\!\pm\! 0.2\%$, measured in the hard state of Cyg X-1 \citep{Krawczynski22}. 

On the other hand, Cyg X-1 is the only source found so far in which the soft X-ray (thermal) reverberation is very weak \citep{Basak25}. Those authors contrast this source with the transient XRB MAXI J1820+070, which exhibits pronounced soft X-ray lags and has a geometry consistent with disk truncation \citep{DeMarco21, Dzielak21, Zdziarski21b}. Soft X-ray lags were also observed in GX 339--4 \citep{DeMarco15var, DeMarco17} and H1743--322 \citep{DeMarco16reverber}. 

To explain this difference, \citet{Basak25} noted that Cyg X-1's relatively low luminosity places it on the lower branch of the hardness-luminosity diagram (soft-to-hard transition for transients), whereas most evidence for disk truncation lies on the upper branch. However, reverberation lags on the lower branch of GX 339--4 were, on average, longer than those on the upper (hard-to-soft) branch \citep{DeMarco17}. We propose that the specific geometry of Cyg X-1 may be related to its accretion mode from the stellar wind. This may continuously replenish the inner parts of the accretion flow, thereby preventing truncation. 

We also note that the low disk temperature we obtained, $0.15\pm 0.01$ keV, implies that disk emission in the 0.5--1 keV range (used to measure the soft lags by \citealt{Basak25}) accounts for only $\approx$1/3 of the total absorbed flux in that range (using the model with a disk and two {\tt reflkerrD} components, which yields nearly the same fraction whether in photons or in energy). Therefore, the intrinsic lag is significantly longer; see figure 20 in \citet{Uttley14}. 

\section*{Acknowledgements}
We thank Andrzej Nied{\'z}wiecki and Phil Uttley for valuable comments. The comments of the referee have allowed us to significantly improve the clarity of this paper. AAZ and MS acknowledge support from the Polish National Science Center through grants 2023/48/Q/ST9/00138 and 2023/50/A/ST9/00527, respectively. SC acknowledges support from the National Science and Technology Council (NSTC) of the Republic of China (Taiwan) under grants NSTC 113--2112--M--007--020 and NSTC 114--2112--M--007--042, and from the Taiwan Space Agency (TASA) under grants NSPO--P--109221 and TASA--P--1140072. BDM acknowledges support from the Spanish MINECO grants PID2023-148661NB-I00, PID2022-136828NB-C44, and CNS2025-166335. This research was also supported by the International Space Science Institute (ISSI) in Bern through the ISSI International Team project `What are the spins of stellar-mass black holes?' (ISSI Team project \#25-660).

\appendix
\section{Calibration issues}
\label{cross}

\begin{figure}
\centerline{\includegraphics[width=1\columnwidth]{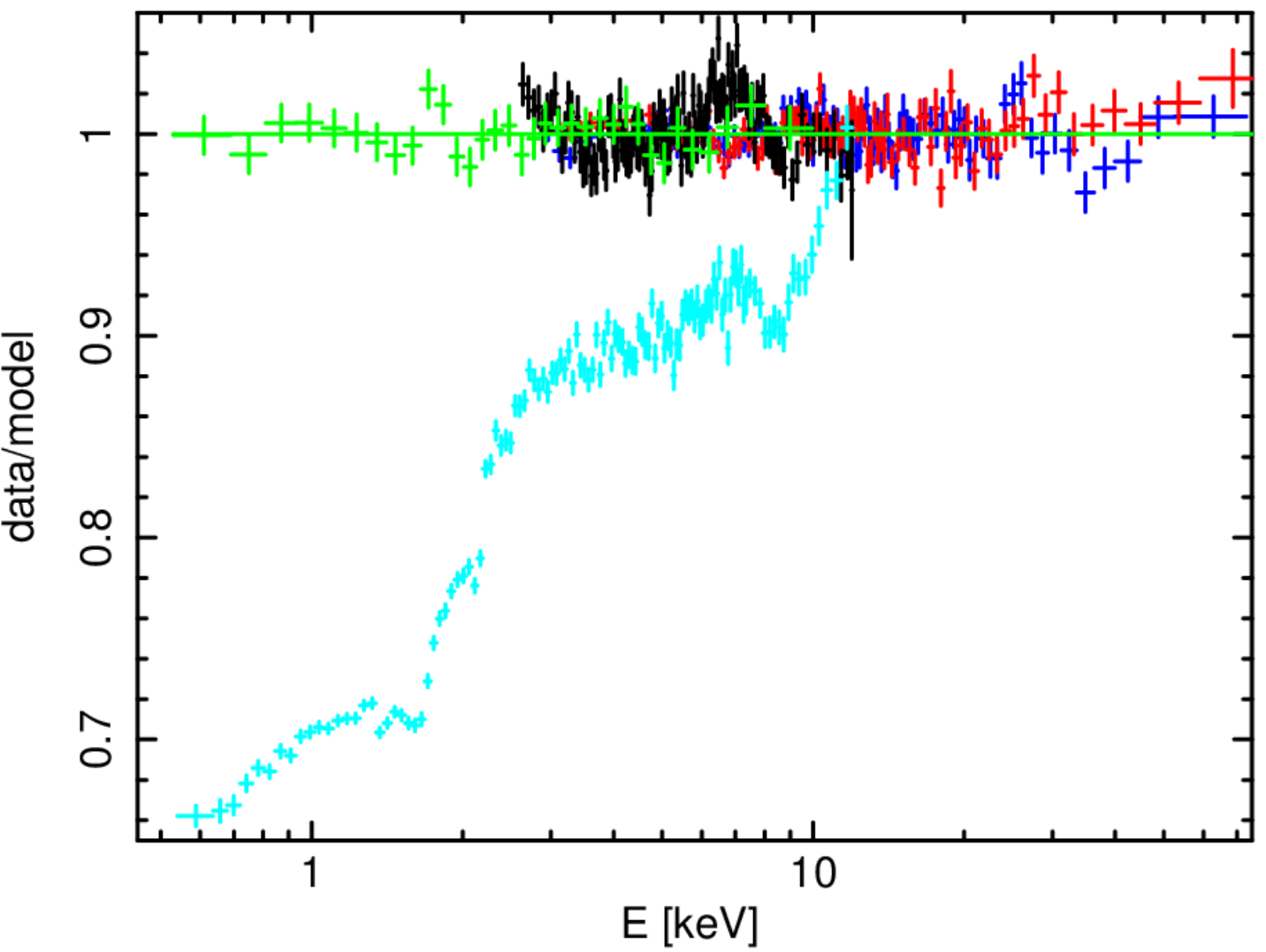}}
\caption{The NICER (green), Resolve (black), and NuSTAR (blue and red) data-to-model ratios for the fit with the two-{\tt reflkerrD} model, covering the full NuSTAR spectral range, 3--79 keV. We observe a substantial excess of Resolve counts in the Fe K range compared with NuSTAR and NICER. In cyan symbols, we also show the Xtend 0.5--10 keV spectrum unfolded with that model (not fitted).
}\label{full}
\end{figure}

In our fits in Section \ref{broad}, we have not included NuSTAR data below 9 keV, and the Fe K region has been fitted primarily with Resolve data. We now fit the data using the full NuSTAR spectral range, 3--79 keV. In this case, statistics in the Fe K region are dominated by NuSTAR. As before, we include the NICER data. We have obtained $\chi^2=4371/4046$ with parameters mostly similar to those obtained when excluding the 3--9 keV NuSTAR data. In particular, the disk can be moderately truncated ($R_{\rm in}\lesssim 10 R_{\rm g}$), but this only slightly improves the fit. The main differences are that the spin parameter is now constrained to be low, $a_*=0.10_{-0.10}^{+0.19}$, and the irradiation index is compatible with 3, $q_1=3.6_{-0.6}^{+0.5}$. However, the shape of the Fe K complex differs between NuSTAR/NICER and Resolve, with the Resolve residuals exhibiting a distinct hump, as shown in Figure \ref{full}. In principle, this could be due to the shorter NuSTAR and NICER exposures and source variability. However, the hardness ratios of both NuSTAR and Resolve remained approximately constant throughout the observation, as shown in Figure \ref{lc}. Thus, this appears to be due to a significant difference in spectral calibration between the instruments, which should be resolved in the future. We also find that the Resolve spectrum increases steeply below the fitted 2.6--12 keV range, whereas the instrument team recommends that data above 1.7 keV be usable.

Figure \ref{full} also shows the unfolded 0.5--10 keV spectrum from the XRISM Xtend instrument. We added that spectrum to the previous fit and found that its shape differs significantly from those of the other detectors. We attempted to fit it, but the fit was very poor, with $\chi^2=5956/4200$ and strong wavy residuals. This indicates an issue with the current Xtend continuum calibration. In Section \ref{data}, we show that the observation mode used resulted in a short effective exposure and short GTIs. It is unclear whether this is related to the spectral discrepancy. We show the light curve used in Figure \ref{hr_flux}; it covers the entire observation interval evenly, indicating that this problem is not related to spectral variability during that interval. A discrepancy between the Xtend and Resolve spectra has also been reported for the BH XRB GS 1354--64 \citep{Liu26}; see also \citet{XRISM25} and \citet{Brenneman25}.  

\bibliographystyle{aasjournal}
\bibliography{allbib} 

\end{document}